\documentclass[useAMS,usenatbib]{mn2e}

\def\apss{Ap\& SS}
\def\apj{ApJ}
\def\apjl{ApJL}
\def\aap{A\& A}
\def\nat{Nature}
\def\araa{ARA\&A}
\def\mnras{MNRAS}

\def\apjs{ApJS}
\def\mnras{{MNRAS}}

\def\'#1{\ifx#1i{\accent"13\i}\else{\accent"13#1}\fi}
\def\alamenos#1{$^{-#1}$}

\def\kms{km sec\alamenos 1}
\def\prom#1{$\langle #1\rangle$}
\def\prommath#1{\langle #1\rangle}

\def\x{{\bf x}}

\def\tauintmath{{\tau_{\rm int}}}
\def\taukinmath{{\tau_{\rm kin}}}

\def\xp{$\tauintmath$}

\title[Six Myths on the Virial Theorem for Interstellar Clouds]{Six Myths on the Virial Theorem for Interstellar Clouds} 

\author[Javier Ballesteros-Paredes]{Javier Ballesteros-Paredes\thanks{E-mail: javier@astrosmo.unam.mx}\\ 
Centro de Radioastronom\'ia y Astrof\'isica, UNAM. Apdo. Postal 72-3
(Xangari), Morelia, Michoc\'an 58089, M\'exico }

\begin{document}


\date{Submitted to MNRAS, \today}

\pagerange{\pageref{firstpage}--\pageref{lastpage}} \pubyear{2006}

\maketitle

\label{firstpage}

\begin{abstract}

The interstellar medium is highly dynamic and turbulent.  However,
little or no attention has been paid in the literature to the
implications that this fact has on the validity of at least six common
assumptions on the Virial Theorem (VT), which are: (i) the only role
of turbulent motions within a cloud is to provide support against
collapse, (ii) the surface terms are negligible compared to the
volumetric ones, (iii) the gravitational term is a binding source for
the clouds since it can be approximated by the gravitational energy,
(iv) the sign of the second-time derivative of the moment of inertia
determines whether the cloud is contracting ($\ddot{I}<0$) or
expanding ($\ddot{I}>0$), (v) interstellar clouds are in Virial
Equilibrium (VE), and (vi) Larson's (1981) relations (mean
density-size and velocity dispersion-size) are the observational proof
that clouds are in VE.  Turbulent, supersonic interstellar clouds
cannot fulfill these assumptions, however, because turbulent
fragmentation will induce flux of mass, moment and energy between the
clouds and their environment, and will favor local collapse while may
disrupt the clouds within a dynamical timescale. It is argued that,
although the observational and numerical evidence suggests that
interstellar clouds are not in VE, the so-called ``Virial Mass''
estimations, which actually should be called ``energy-equipartition
mass'' estimations, are good order-of magnitude estimations of the
actual mass of the clouds just because observational surveys will tend
to detect interstellar clouds appearing to be close to energy
equipartition.  Similarly, order of magnitude estimations of the
energy content of the clouds are reasonable.  However, since clouds
are actually out of VE, as suggested by asymmetrical line profiles,
they should be transient entities.  These results are compatible with
observationally-based estimations for rapid star formation, and call
into question the models for the star formation efficiency based on
clouds being in VE.

\end{abstract}


\section{Introduction}\label{sec:intro}

Interstellar clouds are thought to be turbulent and supersonic.  Their
Mach numbers range from a 1 ($T \sim$ 7000--8000~K, H~I clouds)
through 10 ($T\sim$~10~K, molecular clouds forming low-mass stars) to
50 ($T\sim$~10--50~K, molecular clouds forming high-mass stars).
Since supersonic turbulent motions carry mass and produce
large-amplitude density fluctuations, turbulent
fragmentation\footnote{Turbulent fragmentation is defined as the
process through which a chaotic velocity field produces a clumpy
density structure in the gas within a few dynamical timescales
\cite[see e.g.,][]{VonWeiczacker51, Sasao73, Scalo88, Padoan95, BVS99,
KHM00, HMK01, BP04M}.} is expected to occur in the interstellar
medium.

A useful tool for describing the overall structure of interstellar
clouds is the scalar Virial Theorem (VT).  It is obtained by dotting
the momentum equation by the position vector and integrating over the
volume of interest (\S\ref{sec:VT}).  Being directly derived from the
momentum equation, the VT always holds for any parcel of fluid.

Virial Equilibrium (VE) is a restrictive condition of the VT, and it
is defined by the condition that the parcel of fluid under study has a
second time derivative of the moment of inertia equal to zero.  It has
been invoked extensively to analyze the stability of interstellar
clouds.  However, since in a trans- or super-sonic turbulent
interstellar medium, clouds should be redistributing their mass as a
consequence of their own turbulent motions, it already seems difficult
to achieve VE in a supersonic, turbulent cloud.

Other simplifications, such as the assumption that the cloud is
isolated, or that the surface terms in the VT are negligible, are
frequently made in many (if not in most) astrophysical studies of the
VT.  Those simplifications have been thought to be applicable to
molecular clouds and their substructure for nearly three decades
\cite[e.g., the textbooks by][ and references therein]{Spitzer78,
Shu91, Stahler_Palla05, Lequeux05}, maybe as a consequence of the old
idea that in the ISM, ``all forces are in balance and the medium is
motionless, with no net acceleration'' \cite[][ Chap. 11]{Spitzer78},
in which ``the observational evidence'' seemed to be consistent with
the expectation that interstellar ``clouds tend toward pressure
equilibrium'' \cite[][]{Spitzer78}.

The possible inapplicability of these assumptions has been mentioned
in passing in some previous papers \cite[e.g., ][]{BVS99,
Shadmehri_etal02, BP04P} but no attention has been paid in general to
its implications.  Thus, in the present paper I discuss in detail the
applicability of those assumptions for molecular clouds and their
cores.  In \S\ref{sec:VT} I write explicitly the VT for fluids in its
Lagrangian and Eulerian forms.  In \S\ref{sec:myths} I discuss the six
more common assumptions of the VT and their validity in a turbulent
environment.  In \S\ref{sec:reloading} I explain why even though
clouds are not in VE, they appear to be in energy equipartition, and
argue that asymmetries in the line profiles are the evidence for
clouds out of VE.  Finally, in \S\ref{sec:conclusions} I draw the main
conclusions.

\section{The Virial Theorem}\label{sec:VT}

The Virial Theorem can be derived from the momentum equation, by
dotting it by the position vector $\x$ and integrating it over the
volume of interest.  Althoughit is usually written in its Lagrangian
form, i.e., by following the mass \cite[see, e.g.,][]{Spitzer78,
Shu91, Hartmann98}, it can be also obtained in its Eulerian form,
i.e., by fixing the volume in space \cite[see, e.g.,][]{Parker79,
MZ92}, obtaining

\begin{equation}
{1\over 2} {d^2 I_E\over dt^2} = 
	2\biggl( E_{\rm kin} + E_{\rm int} \biggr) -	
	2\biggl(\taukinmath + \tauintmath \biggr) + M + \tau_M - W 
        -{1\over 2} {d\Phi\over dt}
\label{eq:EVT}
\end{equation}

\begin{equation}
{1\over 2} {d^2 I_L\over d t^2} = 2\biggl(E_{\rm kin} + E_{\rm int}
\biggr) - 2 \tauintmath + M + \tau_M - W 
\label{eq:LVT}
\end{equation}
where $I=\int_V \rho r^2 dV$ is the moment of inertia of the cloud
(subindexes $E$ and $L$ in eqs. [\ref{eq:EVT}] and [\ref{eq:LVT}]
stand for Eulerian and Lagrangian, respectively), $E_{\rm kin} =
1/2\int_V \rho u^2 dV$ and $\taukinmath = -1/2\oint_S x_i\rho u_i u_j
\hat{n}_j dS$ are the kinetic energy of the cloud and the kinetic
stresses evaluated at the surface of the cloud, respectively, $E_{\rm
int} = 3/2\int_V P dV $ is the internal energy, $\tauintmath = -
1/2\oint_S x_i P \hat{n}_i dS$ is the pressure surface term,
$M=1/8\pi\int_V B^2 dV$ is the magnetic energy, $\tau_M = 1/4\pi \oint
x_i B_i B_j \hat{n}_j dS$ is the magnetic stress at the surface of the
cloud, $W=\int_V \ x_i \rho\
\partial \phi / \partial x_i \ dV$ is the gravitational term, with
$\phi$ being the gravitational potential, and $\Phi = \oint_S \rho u_i
r^2 \hat{n}_i dS$ is the flux of moment of inertia through the surface
of the cloud.  In the previous equations, $\rho$, $u_i$, $B_i$, $P$,
and $\hat{n}$ are the density, the $i^{th}$ component of the velocity
$u$, the $i^{th}$ component of the magnetic field $B$, the pressure,
and a unitary vector perpendicular to the surface $S$ that surrounds
the volume $V$, over which the integrals are performed, respectively.
In the notation above it is used the Einstein convention, where
repeated indexes are summed.

%

\section{The common assumptions}\label{sec:myths}

Various assumptions for the terms involved in eqs. (\ref{eq:EVT}) and
(\ref{eq:LVT}) have been made in the literature.  Some of them,
indeed, have been converted into myths, since their applicability to
interstellar clouds not only is not demonstrated, but it is not even
questioned, either in textbooks, or research papers.

\subsection{First assumption: The kinetic energy is generally a term
  of support} {\label{sec:support}}

It is almost universally considered in the literature that the
turbulent (or kinetic) energy, $E_{\rm kin} = 1/2 \int \rho u^2\ dV$,
provides support to clouds against collapse.  While this is true for a
system of particles, and partially valid if the kinetic energy is in
the form of large-scale expansion and/or rotation, it is by no means
certain that all the kinetic energy available will help against
collapse in a system where turbulent fragmentation can occur, as
discussed below.

This idea has its origins in \citet{Chandra51b}, who proposed that in
the analysis of the gravitational instability the turbulent velocity
field should be included.  In his description, an effective sound
speed is introduced, given by

\begin{equation}
c^2_{\rm eff} = c_s^2 + {1\over 3} u^2_{\rm rms}
\label{eq:cs_eff}
\end{equation}
where $c_s$ is the sound speed, and $u_{\rm rms}$ is the velocity
dispersion of the turbulent motions \cite[see, e.g.,][ for a
review]{KHM00, MK04}.  This description is valid only if (a) turbulent
motions are confined to scales much smaller than the size of the cloud
\cite[][]{BVS99}, and (b) such motions do not produce new,
smaller-scale Jeans-unstable density enhancements.  The first
hypothesis disregards one of the main features of turbulent flows in
general \cite[e.g.,][]{Kolmogorov41, Lesieur90}, and of interstellar
clouds in particular \cite[][]{Larson81}, namely, that the largest
velocities occur at the largest scales.  An attempt to include this
fact has been proposed by \citet[][]{Bonazzola_etal87}, who suggested
including the value of the rms velocity dispersion {\it at each scale}
$l\propto 1/k$, i.e.,

\begin{equation}
c^2_{\rm eff}(k) = c_s^2 + {1\over 3} \prommath{u(k)}^2
\label{eq:cs_eff_k}
\end{equation}
where $k$ the wavenumber corresponding to the scale $l$, and
\prom{u(k)} is given by the energy spectrum $E(k) = C k^{-\delta}$ as

\begin{equation}
\prommath{u(k)}^2 = \int_k^\infty E(k) dk =   {C\over 1- \delta} k^{1-\delta}\ .
\label{eq:cs_eff_bona}
\end{equation}
where $C$ is a constant and $\delta$ is the spectral index.  The
second condition, i.e., that turbulent motions do not produce
Jeans-unstable density enhancements, has the underlying complication
that motions at scales larger than $l\sim 1/k$ will be very
anisotropic with respect to structures of size $l$.  Those modes will
produce shear (through vortical modes) or compressions (through
compressible modes) to the structures\footnote{It should be remembered
that compressible and vortical modes are coupled, and they exchange
energy \cite[e.g.,][]{Sasao73, VPP96, Kornreich_Scalo00,
Elmegreen_Scalo04}.}.  Compressions in particular reduce the local
Jeans mass \cite[][]{Sasao73, Hunter_Fleck82}, and can induce local
collapse.  Thus, a fraction of the turbulent kinetic energy is
involved in promoting collapse, rather than opposing it.

By decomposing the velocity field in its solenoidal and compressible
components, the kinetic energy modes that provide support to the
clouds are those having divergence larger or equal to zero,

\begin{equation}
\nabla \cdot u \geq 0.
\label{eq:kinetic_support}
\end{equation}
This includes the solenoidal modes ($\nabla\cdot u = 0$), and the
expansional component of the compressible modes ($\nabla\cdot u > 0$).
In other words, the precise result of collapse or support must then
reflect the balance between all the agents that favor collapse against
those agents that provide support.  In the first group, not only the
gravitational energy should be included, but also the kinetic energy
involved in the compressible modes ($E_{{\rm kin}, \nabla \cdot u <
0}$), versus the kinetic energy involved in the expansional and
rotational modes ($E_{{\rm kin},\nabla\cdot u \geq 0}$).

\subsection{Second assumption: The surface terms are negligible}

It is frequently found in the literature that the surface terms are
neglected altogether, especially in observational work
\cite[][]{Larson81, MG88, Fuller_Myers92}, mainly because there is
not a direct way of measuring them observationally, although it is
also a common practice in theoretical studies \cite[e.g.,
][]{Chandra_Fermi53, Parker69} and textbooks \cite[e.g.,][]{Spitzer78,
Parker79, Shu91, Stahler_Palla05}.  This assumption is based on the
idea that self-gravitating clouds may be considered isolated because
then their internal energies dominate the dynamics.  The most notable
exception is the thermal pressure surface term \xp, which is
frequently invoked for ``pressure confinement''
\cite[e.g.,][]{McCrea57, Keto_Myers86, BM92, Yonekura_etal97}.
Although some works have considered, by analogy, the possibility of
turbulent pressure confinement by means of the term $\taukinmath$
\cite[e.g.,][]{MZ92}, such confinement of a cloud is difficult to
achieve because the large-scale turbulent motions are anisotropic and
will in general distort or even disrupt the cloud \cite[][]{BVS99}.

Although we cannot measure the surface terms from observations, the
possibility that they are as important as their corresponding
volumetric terms suggest to investigate into numerical simulations of
the interstellar medium.  In fact, \citet[][]{BV97} found that for an
ensemble of clouds in two-dimensional simulations of the interstellar
medium at a kiloparsec scale, the surface terms have magnitudes as
large as those of the volumetric ones \cite[for three-dimensional
simulations, see also][]{Shadmehri_etal02, TP04, DVK06}.  This result
suggests that, on one hand, either surface and volumetric terms are of
comparable importance in shaping and supporting the clouds.  On the
other hand, it suggests that clouds must be interchanging mass,
momentum and energy with the surrounding medium.  In such an
environment, the meaning of thermal or ram pressure confinement is not
clear, since motions at all scales must morph and deform the cloud.

 \subsection{Third assumption: The gravitational term is the gravitational
energy}

The gravitational term entering the VT is written as

\begin{equation}
W = - \int_V  x_i\ \rho\ {\partial \phi \over \partial x_i}\ dV.
\label{eq:w}
\end{equation}
Splitting-up the gravitational potential as the contribution from the
cloud itself ($\phi_{\rm cloud}$), plus the contribution from the
outside ($ \phi_{\rm ext}$),

\begin{equation}
\phi = \phi_{\rm cloud} + \phi_{\rm ext}\ ,
\label{eq:phi}
\end{equation}
the gravitational term can be written as:

\begin{equation}
W = E_g - \int_V x_i\ \rho\ {\partial \phi_{\rm ext}\over \partial
             x_i}\ dV\ ,
\label{eq:Eg1_Eg2}
\end{equation}
where $E_g = -1/2 \int_V \rho\phi\ dV$ is the gravitational energy of
the cloud alone, since the volume of integration of $\phi_{\rm cloud}$
and the volume $V$ of the integral coincide
\cite[e.g.,][]{Chandra_Fermi53}.  The second term in the right-hand
side is usually either implicitly or explicitly assumed to be
negligible compared to the first term, and the gravitational term $W$
is then assumed to equal the gravitational energy $E_g$.  This is
valid only if the distribution of mass is spheroidal, or if the medium
outside the cloud is tenuous such that its contribution to the
potential is negligible.  However, clouds are more similar to
irregular fractals with arbitrary shapes (frequently filamentary) than
to spheroids \cite[e.g.,][]{Falgarone_etal91}, and the contribution
from the external gravitational field may not be negligible, giving
rise to tidal torques.  Although up to now there is no observational
estimation of the contribution of the external mass to the
gravitational term for any interstellar cloud,
mass estimates for for H~I ``envelopes'' around molecular clouds are
of the same order of magnitude than the mass of the molecular clouds
themselves \cite[e.g.,][]{Williams_Maddalena96,
Moriarty-Schieven_etal97}.  Thus, it is not difficult to realize that
the contribution of the second term in the right-hand side of
eq. (\ref{eq:Eg1_Eg2}) to the gravitational term $W$ can be of the
same order of magnitude as the gravitational energy $E_g$.

Similar arguments can be made for the interiors of molecular clouds:
even though we can approximate their shapes as (triaxial) spheroids
\cite[][]{Jijina_etal99}, embedded molecular cloud cores are subject not
only to their own self-gravity, but also to the tidal forces from
their parental molecular cloud.  Thus, it is not clear that the tidal
forces represented in the second term of eq.  (\ref{eq:Eg1_Eg2}) will
be negligible, and the assumption that the gravitational term equals
the gravitational energy of the cloud seems unjustified.

What is the meaning and the effect of the second term of eq.
(\ref{eq:Eg1_Eg2}) on the energy budget of the clouds?  It can be seen
that it involves the gradient of the external potential.  For
non-symmetrical distributions of mass, this term is out of balance
even if the distribution of mass inside the cloud is symmetric.  Thus,
this term represents the tidal forces over the mass contained in the
volume $V$, and it can be split up into three terms:

\begin{eqnarray}
 \int_V x_i\ \rho\ {\partial \phi_{\rm ext}
/ \partial x_i}\ dV = 
       \int_V {\partial (\phi_{\rm ext}\ x_i\ \rho)
/ \partial x_i}\ dV - \nonumber \\
       3 \int  \phi_{\rm ext}\ \rho\ dV  - 
        \int  \phi_{\rm ext}\ x_i\ {\partial \rho
/ \partial x_i}\ dV \ .
\label{eq:w-descomposicion}
\end{eqnarray}

Using the Gauss theorem, the first term on the right-hand side of
eq. (\ref{eq:w-descomposicion}) can be interpreted as the
gravitational pressure evaluated at the boundary of the cloud,

\begin{equation}
 \int_V {\partial (\phi_{\rm ext}\ \rho\ x_i) \over \partial x_i}\ dV
 = \oint_S (\phi_{\rm ext}\ \rho\ x_i) \ \hat{n_i}\ dS.
\label{eq:gravitational_pressure}
\end{equation}

The second term, by similarity with the gravitational energy, can be
interpreted as three times the work done to assemble the density
distribution of the cloud against the external mass,

\begin{equation}
E_{g,{\rm ext}} = 3 \int \phi_{\rm ext}\ \rho\ dV.
\end{equation}
Finally, the last term in eq (\ref{eq:w}),

\begin{equation}
 \int \phi_{\rm ext}\ x_i\ {\partial \rho\over \partial x_i}\ dV
\label{eq:grad_rho}
 \end{equation}
involves the gradient of the density field inside the cloud.  Although
there is no clear interpretation of this term, it is worthwhile noting
that its contribution is null for a homogeneous distribution of mass
inside the volume $V$ of integration.

 From the numerical point of view, several studies \cite[][]{BV97,
 BP99, Shadmehri_etal02, TP04, DVK06} have found that the gravitational
 term can be negative or positive.  If negative, it will be a
 confining agent.  If positive, its overall action will be to
 contribute to the disruption of the cloud/core.

 \subsection{Fourth assumption: the sign of $\ddot{I}$ defines whether the
 cloud is collapsing or expanding}

It is frequently argued in the literature that if $\ddot{I} > 0$, the
cloud must be expanding, while $\ddot{I} < 0$ implies that the cloud
is contracting.  Equilibrium is assumed to occur when $\ddot{I}=0$.

This idea rests on the fact that the gravitational energy $E_g$ has a
negative sign, and it is a confining agent, while the internal
energies (thermal, kinetic, or magnetic) are positive, and they are
assumed to act as supporting agents against collapse.  By neglecting
the other terms, it can be assumed that if the gravitational energy is
larger than the sum of the internal energies, the sign of the
right-hand side of eq. (\ref{eq:LVT}) is negative.  Physically, if
gravity wins, the cloud collapses.

Although energetically this is true, it is not hard to find an example
in which an expanding cloud has a negative second time derivative of
the moment of inertia.  Assume a sphere with constant density and
fixed mass $M$.  Its moment of inertia is

\begin{equation}
I = {3\over 5}\pi M R^2.
\end{equation}
If its size varies with time, for instance, as a power law,

\begin{equation}
R = R(t) = R_0 \biggl(  {t\over t_0} \biggr)^\gamma\ ,
\end{equation}
its second time derivative is given by

\begin{equation}
\ddot{I} = {6 M\over 5}\ \biggl( {R_0\over t_0} \biggr)^2 \
              \biggl({t\over t_0} \biggr)^{2\gamma-2}  \
              \gamma\ \biggl( 2\ \gamma - 1 \biggr) \ ,
\end{equation}
which is negative if $0 < \gamma < 1/2$, even though it is expanding.
In general, $\ddot{I}$ has been treated in the literature as if it
were $\dot{I}$.

\subsection{Fifth assumption: interstellar clouds are in Virial Equilibrium} 

The definition of Virial Equilibrium is that the left-hand side of the
Lagrangian Virial Theorem (eq. [\ref{eq:LVT}]) equals zero:

\begin{equation}
\ddot{I}_L=0
\label{eq:VE}
\end{equation}
\cite[see, e.g.,][]{Spitzer78}.  
  Although there are some observational papers showing molecular
  clouds out of VE \cite[e.g.,][]{Carr87, Loren89, Heyer_etal01}, it
  is frequently encountered in the literature the VE assumption for
  MCs and their substructure \cite[e.g.,][]{MG88, McKee99,
  Krumholz_McKee05, Ward-Thompson_etal06, Tan_etal06}.
  As mentioned in the introduction, this statement is based on the
  (old) idea that all the forces (in the ISM) should be in balance,
  and the medium should have no net acceleration \cite[see e.g.,] [,
  Chap.  11]{Spitzer78}.  However, both observational
  \cite[][]{Jenkins_etal83, Bowyer_etal95, Jenkins_Tripp01, Jenkins02,
  Redfield_Linsky04}, and numerical \cite[][]{VS_etal03,
  MacLow_etal05, Gazol_etal05} studies have found that the ISM is not
  in strict pressure balance, but exhibits strong pressure
  fluctuations, and in fact the turbulent ram pressure is
  significantly larger than the thermal one
  \cite[e.g.,][]{Boulares_Cox90}.

  From an observational point of view, it cannot be demonstrated that
  clouds are in VE because neither the detailed three-dimensional
  density structure of molecular clouds, nor the time derivatives for
  interstellar clouds can be measured observationally.  From a
  theoretical point of view, the VE assumption has strong
  implications: either supersonically turbulent clouds are not
  redistributing their mass inside, or the way in which the
  time-derivative of the moment of inertia ($\dot{I}$) vary (which can
  be interpreted as the time-variation of their mass redistribution)
  is constant.  Both statements seem implausible in a highly
  dynamical, nonlinear ISM.  \citet[][]{MZ92}, for instance, have
  recognized the difficulty of achieving VE in a turbulent medium,
  because turbulent motions carry fluid elements, redistributing their
  mass.

  From the numerical point of view, fortunately, it is possible to
  calculate all the terms of the VT for clouds, since we know all the
  variables involved in the numerical simulations.  In order to test
  the Virial theorem, \citet[][ see also \citealt{BVS99, BP99}]{BV97}
  calculated all the integrals in the EVT for clouds in numerical
  simulations of the interstellar medium at a kiloparsec scale by
  \citet{PVP95}.  They found that the second time derivative of the
  moment of inertia never goes to zero, suggesting that clouds must be
  transient entities.

 \citet{McKee99} has recognized the difficulty to achieving VE for
 actual MCs.  He proposed two possibilities for assuming clouds in VE:
 the first one is to average in time the second time derivative of the
 moment of inertia.  He suggests that $\prommath{\ddot I}=0$ if the
 averaging time considered is much larger than the dynamical timescale
 of the cloud, i.e., if $t_{\rm avg} \gg t_{\rm dyn}$.  The second one
 is that for an ensemble of clouds, some of them may have positive
 values of $\ddot I$, and some others will have negative values, so
 that VE holds for the ensemble.  Regarding the first assumption,
 there is a hidden assumption behind it: the cloud maintains its
 identity during several dynamical timescales\footnote{In fact, it has
 to be oscillating around a mean shape without a strong redistribution
 of mass, in order to achieve the condition $\prommath{\ddot I}=0$.  A
 similar assumption is made by \citet[][]{MZ92}.}.  However, from the
 numerical point of view, simulations of the ISM suggest that the
 clouds are continually morphing and exchanging mass, energy and
 momentum with their surroundings, so that over $t\sim t_{\rm dyn}$
 they have changed significantly.  On the other hand, observational
 evidence \cite[][]{BHV99, Elmegreen00, HBB01, BH06} also suggests
 that the lifetimes of the clouds are not significantly larger than
 their own dynamical times (i.e., they are transient).  Concerning the
 second assumption, the analysis by \citet[][ see also
 \citealt{BP99}]{BV97} shows that the moment of inertia spans up to 7
 orders of magnitude (in absolute value) between the largest clouds
 and the smallest. Thus, it is not clear that an average for such
 large scatter will be representative of the actual dynamics of the
 clouds.  In other words, even if $\ddot{I}$ did average out to zero
 for the ensemble, this does not alter the fact that, individually,
 clouds are not in VE.

\subsection{Sixth assumption: Larson's relationships are the observational
 demonstration of clouds being in Virial Equilibrium}

The mean density-size and velocity dispersion-size relations first
discussed by \citet{Larson81} are thought to be an observational
demonstration of Virial Equilibrium.  However, as \citet{MG88}
recognize, they are actually only compatible with equipartition.  For
instance, assuming equipartition between kinetic (turbulent) and
gravitational energy,

\begin{equation}
\delta v^2 \sim {GM\over R} \propto \prommath{\rho} R^2.
\label{eq:equipartition}
\end{equation}
Thus, if there is a mean density-size power-law relationship $\rho
\propto R^\alpha$, then, there should be a velocity dispersion-size
relationship of the form

\begin{equation}
\delta v^2 \propto R^\beta ,
\end{equation}
with 

\begin{equation}
\beta = (\alpha + 2)/2 .
\label{eq:beta-alpha}
\end{equation}
Several caveats must be mentioned at this point.  First of all, this
derivation does not mean VE, but energy equipartition, since the only
assumption was $E_g\sim E_{\rm kin}$ (see
eq. [\ref{eq:equipartition}]).  A similar result can be found if the
equipartition assumed is valid between the gravitational and magnetic
energy.  In this case, $\delta v $ is proportional to the Alfv\'en
speed \cite[see, e.g.,][]{MG88}.
 Second, as discussed by \citet{VSG95}, in the particular case of
 $\rho \propto R^{-1}$, $\delta v \propto R^{1/2}$.  However, the pair
 $\alpha = -1$, $\beta = 1/2$ is not unique.  Any pair of values
 satisfying the equation (\ref{eq:beta-alpha}) will be consistent with
 energy equipartition (again, not VE).
Finally, it is convenient to recall that the validity of Larson's
relations has been called into question, especially the mean
density-size relationship \cite[][]{Kegel89, Scalo90, VBR97, BM02}.
It seems that this relationship is more a consequence of the
observational process, in which the dynamical range of the
observations is limited below by the minimum sensitivity of the
telescopes, and above by saturation of the detectors, optically thick
effects, and depletion.

 \section{Discussion: Virial Mass, Virial Equilibrium and the Star
 Formation Efficiency}\label{sec:reloading}

As it is discussed, there are at least six common assumptions related
to the Virial Theorem which seem to be unjustified for a turbulent
ISM.  Some questions, such as why Virial Masses are good
order-of-magnitude estimations of the actual mass, or why in
principle, the sub- or super-criticality of a molecular cloud core is
a good estimation of the dynamical state of such a core, are still
valid to ask.

The answer is that those estimations are based on energy
equipartition, not on VE (i.e., $E_{\rm kin} \sim E_g \sim E_M$ does
not mean that $\ddot{I}_L$ is negligible when it is compared to the
left-hand side of eq. [\ref{eq:LVT}]).  Molecular clouds seem to be in
approximated equipartition between self-gravity, kinetic, and magnetic
energy \cite[see, e.g.,][]{MG88} although the super-Alfv\'enic nature
of molecular clouds is still a matter of debate, \cite[e.g., ][ see
also \citealt{Ward-Thompson_etal06} for an observational review;
\citealt{BKMV06}, for a theoretical one]{PN99}.

The question thus, is why MCs seem to be in energy equipartition?  The
answer is related to what we identify as a cloud, and to observational
limitations: in the first case, clouds with a substantial excess of
internal energy will rapidly expand and merge with the more diffuse
medium.  This is the case of an H~II region or SN explosion expanding
within its parental molecular cloud.  The amount of energy provided by
those events is enough to disrupt their parental environment within a
few Myr \cite[e.g.,][]{Franco_etal94}.  The excess of internal energy
is mostly in the form of UV photons, which rapidly ionize the
molecular gas, reducing even more the life timescale of what it is
identified as the parental molecular cloud \cite[see,
e.g.,][]{Mellema_etal06}.
In the second case, if the cloud has an excess of gravitational
energy, its free-fall time velocity is at most a factor of $\sqrt{2}$
the velocity dispersion needed for equipartition.  This means that the
difference between a free-fall collapsing cloud and one in equilibrium
is just a factor of two in the energy, and a factor of $\sqrt{2}$ in
velocity.  Both systems will be, in principle, in order-of-magnitude
energy equipartition.  Note, furthermore, that in the free-fall
collapsing case all the kinetic energy will be observationally
identified as energy for support, where it is not actually providing
any support.  This re-enforces the need to distinguish between the
compressible, expansive and vortical modes of the kinetic energy (see
\S\ref{sec:support}).

As discussed above, not all the terms entering the VT can be measured
observationally.  However, all of them can be measured from numerical
simulations.  The work by \cite{BV95, BV97, Shadmehri_etal02, TP04}
for turbulent realizations of MCs shows that all the terms entering
the VT are of similar importance.  Note that the same problems related
to the identification of an expanding cloud are applicable to the
numerical work.  In this case, what we identify as a cloud are the
density enhancements.  By definition, vacuumed regions (due either to
modeled stellar activity, or by the turbulence itself) are no longer
considered clouds.  As for the collapsing case, strongly
self-gravitating clouds will develop a velocity field that is a factor
of $\sqrt{2}$ the value of energy equipartition, if only gravity is
acting, and of the order of magnitude if the turbulence is forced.
%
Thus, order of magnitude equipartition between magnetic, kinetic, and
gravitational energy is valid to assume to predict whether the clouds
should be supercritical or subcritical \cite[][]{BM92, Nakano98}, or
to understand why subcritical cores are unlikely to survive
long-lifetimes within MCs \cite[][]{Nakano98}.  However, the
order-of-magnitude coincidence of the involved energies does not mean
that clouds are in equilibrium at all.  It should be stressed that the
difference is not just semantic (Virial mass vs energy-equipartition
mass).  The difference is conceptual: turbulent clouds cannot be in
equilibrium because turbulent fragmentation takes them out of
equilibrium, and thus they do not last long.  In order to achieve
equilibrium, (even if it is a time-averaged equilibrium,
$\prommath{\ddot I}=0$, as suggested by \citet{McKee99}), it is
necessary to allow the cloud to live for, at least, several crossing
times.  To give an example, the Taurus Molecular Cloud has a size of
$l \sim 20$ pc, and a velocity dispersion of $\delta v\sim 2$~\kms.
Its lateral crossing time is $\tau_{\rm cross} = {l/ \delta v} \sim
10\ {\rm Myr}$.  In order to achieve equilibrium, Taurus has to live
20-30 Myr without a serious distortion or modification.  Such a
condition seems difficult to achieve if one realizes that Taurus has a
Mach number of the order of 10, and the associated H~I gas has Mach
numbers of 20-30 in respect to the internal CO gas \cite[see, e.g.,
the velocity-position diagrams of H~I and CO in][]{BHV99}.

  The fact that turbulent clouds cannot be in VE and cannot last
  several crossing times is in clear contradiction with
  \citet[][]{Tan_etal06}, who made theoretical arguments that favor
  star formation occurring during several dynamical timescales.
  However, in their theoretical derivation they applied the star
  formation rate per free-fall time expression given by eq. (30) of
  \citet[][]{Krumholz_McKee05} to a clump in hydrostatic equilibrium.
  As \citet{Krumholz_McKee05} point out, their eq. (30) is valid only
  for large Mach numbers\footnote{Other problems involved in the
  derivation of eq (30) by \citet[][]{Krumholz_McKee05} will be
  discussed in a further contribution.} (20-40), while by definition,
  a clump in hydrostatic equilibrium is subsonic.  Thus, the
  theoretical arguments in favor of slow star formation are wrong

Another point to be stressed concerns line profiles of H~I, CO and its
isotopes, and even higher density tracers (CS, NH$_3$, etc.).  If
interstellar clouds and/or their cores are not in Virial Equilibrium,
they must be distorted and disrupted within a dynamical timescale.
How will their line-profiles look, and how will they look if they were
in equilibrium?  Certainly, if turbulence were microscopic (necessary
to achieve equilibrium), line profiles should be symmetric.  The fact
that H~I and CO clouds \cite[e.g., ][]{Hartmann_Burton97, Dame_etal01}
as well as molecular cloud cores \cite[e.g., ][]{Falgarone_etal98}
exhibit non-symmetric line profiles is probably the best and the only
evidence that the term $\ddot{I}_L$ is different from
zero\footnote{Note that in an Eulerian frame of reference, an
idealized dustlane in a spiral wave may appear to be long-lived in
terms of its overall structure even though particular gas molecules do
not stay inside it very long.  For such system the line profile will
be asymmetric, while its Eulerian second-time derivative of the moment
of inertia can be zero ($\ddot{I}_E = 0$).  This system, however, is
not in VE, as the definition of VE, eq. (\ref{eq:VE}) involves the
{\it Lagrangian} frame of reference \cite[Note that the difference
between the Eulerian and Lagrangian VT is $\ddot{I}_L- \ddot{I}_E =
-4\tau_{\rm kin} + d\Phi/dt$, as pointed out by][]{MZ92}.}.
Even a small degree of asymmetry in the line profiles suggests that
large-scale motions are present in the observed system, which is a
natural consequence of turbulence being a multiscale phenomenon.  This
does not mean that such a system is collapsing or expanding as a
whole, as formerly suggested by \citet[][]{Goldreich_Kwan74}.  Instead,
it means that turbulent large-scale motions are present in the system,
which should be evolving within a dynamical timescale.

In this context, it should be stressed that clouds presenting
large-scale motions and evolving to form stars rapidly not necessarily
will have a high star formation efficiency, as is the common belief
since \cite{Zuckerman_Evans74}.  Although former models of
quasi-static evolution of molecular clouds were proposed to reduce the
star formation efficiency, the turbulent models producing local
collapse rapidly, have small efficiency because gravo-turbulent
fragmentation involves only a small fraction of the mass of the system
in collapsing regions \cite[][]{VBK03, VKB05}.  In numerical models,
when energy feedback from stars is included, the cloud is blowed out
rapidly \cite[see, e.g.,][]{BP04P}.  However, it should be recognized
that a detailed quantification of the star formation efficiency in
turbulent simulations with open boundary conditions and stellar
feedback is needed.

\section{Conclusions}\label{sec:conclusions}

The present contribution has discussed the applicability of the six
more common assumptions on the Virial Theorem.  Specifically,

\begin{enumerate}

 \item{} It was shown that a decomposition of the velocity field into
 its vortical and compressible modes is necessary, since only modes
 satisfying the condition $\nabla\cdot u > 0$ provide support, while
 modes satisfying $\nabla\cdot u < 0$ foment collapse.

 \item{} It was argued that for a supersonic, turbulent ISM, surface
 terms should not be neglected.

 \item{} It was shown that the gravitational term can be decomposed
 into a contribution from the cloud itself, and a contribution from
 the outside.  The first part is the well-known gravitational energy.
 The second part is the sum of three terms: The gravitational pressure
 evaluated at the surface of the cloud plus three times the work done
 against the external mass to assemble the density distribution of the
 cloud, plus a term that depends on the gradient of the density
 distribution of the cloud.  These represent the tidal forces due to
 an external potential, as could be the case of a dense core within a
 giant molecular cloud, or a giant molecular cloud close to a spiral
 arm.  It is argued that this contribution can be as important as the
 self-gravitational energy.

 \item{} Using a simple counter-example, it was shown that the sign of
 the second-time derivative of the moment of inertia does not
 determine whether the cloud is contracting or expanding. An expanding
 cloud may very well satisfy the condition $\ddot{I}<0$, and a
 contracting one may satisfy $\ddot{I}>0$, contrary to the common
 belief.  In other words, $\ddot{I}$ has been treated in the
 literature as if it were $\dot{I}$.

 \item{} It was argued that interstellar clouds are not likely to
 satisfy the Virial Equilibrium (VE) condition $\ddot{I}=0$.

 \item{} It was shown that Larson's (1981) relations are not
 observational proof for clouds being in VE.

 \item{} Clouds seem to be in energy equipartition because of either
 observational limitations, as well as because of the intrinsic
 definition of a cloud.

\end{enumerate}

Turbulent fragmentation plays a crucial role for the inapplicability
of the VT to interstellar clouds, since it will induce a flux of mass,
moment and energy between the clouds and their environment, and will
favor local collapse while disrupting the clouds within a dynamical
timescale.  The common assumptions discussed in the present
contribution drive our understanding of the dynamical state of the
interstellar clouds toward a picture that favors a static ISM.
However, they are highly difficult to fulfill if the ISM is highly
turbulent, as it was found to be many years ago \cite[e.g.,
][]{McCray_Snow79}.  Inferences of the star formation efficiency for
supersonic (Mach numbers $\sim 20-40$) clouds in virial equilibrium
living several dynamical timescales \citep[e.g., ][]{Krumholz_McKee05,
Tan_etal06} should be taken with caution.

The lack of observational evidence for clouds being in VE, and the
identificaion of asymmetrical line profiles observed toward
interstellar clouds using different tracers are the best evidence of
clouds being out-of-equilibrium systems.
These facts lead us to the conclusion that clouds should be transient
structures, which exchange mass, momentum and energy with their
environment.  This is precisely the opposite point of view to the old
one in which clouds should be at rest, which still is present in
textbooks and papers, but it is consistent with recent
observationally-based works that favor a scenario of rapid cloud and
star formation \cite[see, e.g., the reviews by ][ and references
therein.]{MK04, BKMV06}.

\section*{Acknowledgements}
I want to thank Paola D'Alessio, and Hebe Vessuri who encouraged me to
write this paper.  To Paola D'Alessio, Lee Hartmann, Ralf Klessen,
Stan Kurtz, Mordecai-M. Mac Low, Padelis Papadopoulos and Enrique
V\'azquez-Semadeni, for the careful reading of this manuscript and
their useful comments.  Special thanks to Lee and Enrique for fruitful
and enjoyable discussions on the subject, and to Bruce Elmegreen, the
referee, for an encouraging referee report.  Thanks also to Mary-Ann
Hall (mahall@mail.unla.edu.mx) for proofreading this paper.  This work
was supported by UNAM-PAPIIT grant number 110606, and made extensive
use of the NASA-ADS database.

\label{lastpage}


\begin{thebibliography}{}


\bibitem[Ballesteros Paredes(1999)]{BP99} Ballesteros Paredes, J.\
1999, Ph.D.~Thesis,

\bibitem[Ballesteros-Paredes(2004a)]{BP04P}  Ballesteros-Paredes, J.\
2004, \apss, 289, 243  

\bibitem[Ballesteros-Paredes(2004b)]{BP04M} Ballesteros-Paredes, J.\
2004, \apss, 292, 193

\bibitem[Ballesteros-Paredes \& Hartmann(2006)]{BH06}
Ballesteros-Paredes, J., \& Hartmann, L.\ 2006.
Rev. Mex. Astron. Astrophys, submitted ({\tt astro-ph/0605268})



\bibitem[Ballesteros-Paredes et al.(1999)]{BHV99} Ballesteros-Paredes,
J., Hartmann, L., \& V{\'a}zquez-Semadeni, E.\ 1999, \apj, 527, 285

\bibitem[Ballesteros-Paredes et al.(2006)]{BKMV06}
Ballesteros-Paredes, J., Klessen, R.~S., Mac Low, M.~-., \&
Vazquez-Semadeni, E.\ 2006, in Protostars and Planets V.  ({\tt
astro-ph/0603357})

\bibitem[Ballesteros-Paredes \& Mac Low(2002)]{BM02}
Ballesteros-Paredes, J., \& Mac Low, M.-M.\ 2002, \apj, 570, 734

\bibitem[Ballesteros-Paredes \& V\'azquez-Semadeni(1995)]{BV95}
Ballesteros-Paredes, J., \& V\'azquez-Semadeni, E.\ 1995, Revista
Mexicana de Astronom\'ia y Astrof\'isica, Conf. Series, 3, 105

\bibitem[Ballesteros-Paredes \& V{\'a}zquez-Semadeni(1997)]{BV97}
Ballesteros-Paredes, J., \& V{\'a}zquez-Semadeni, E.\ 1997, American
Institute of Physics Conference Series, 393, 81

\bibitem[Ballesteros-Paredes et~al.(1999a)]{BVS99}
{Ballesteros-Paredes}, J., {V\'azquez-Semadeni}, E., and {Scalo},
J. 1999a, \apj, 515, 286--303

\bibitem[Bertoldi \& McKee(1992)]{BM92} Bertoldi, F., \& McKee, C.~F.\
1992, \apj, 395, 140

\bibitem[Bonazzola et al.(1987)]{Bonazzola_etal87} Bonazzola, S.,
Heyvaerts, J., Falgarone, E., Perault, M., \& Puget, J.~L.\ 1987,
\aap, 172, 293

\bibitem[Boulares \& Cox(1990)]{Boulares_Cox90} Boulares, A., \& Cox,
D.~P.\ 1990, \apj, 365, 544

 \bibitem[Bowyer et al.(1995)]{Bowyer_etal95} Bowyer, S., Lieu, R.,
 Sidher, S.~D., Lampton, M., \& Knude, J.\ 1995, \nat, 375, 212
 

\bibitem[Carr(1987)]{Carr87} Carr, J.~S.\ 1987, \apj, 323, 170 

\bibitem[Chandrasekhar(1951)]{Chandra51b} Chandrasekhar, S. (1951)
Proc. R. Soc. London A, 210, 26

\bibitem[Chandrasekhar \& Fermi(1953)]{Chandra_Fermi53} Chandrasekhar,
S., \& Fermi, E.\ 1953, \apj, 118, 116

\bibitem[Dame et al.(2001)]{Dame_etal01} Dame, T.~M., Hartmann, D., \&
Thaddeus, P.\ 2001, \apj, 547, 792


\bibitem[Dib et al.(2006)]{DVK06} Dib, S., et al. 2006, in preparation

\bibitem[Elmegreen(2000)]{Elmegreen00} Elmegreen, B.~G.\ 2000, \apj,
530, 277


\bibitem[Elmegreen \& Scalo(2004)]{Elmegreen_Scalo04} Elmegreen, B.~G., 
\& Scalo, J.\ 2004, \araa, 42, 211 

\bibitem[Falgarone et al.(1998)]{Falgarone_etal98} Falgarone, E.,
Panis, J.-F., Heithausen, A., Perault, M., Stutzki, J., Puget, J.-L.,
\& Bensch, F.\ 1998, \aap, 331, 669

\bibitem[Falgarone et al.(1991)]{Falgarone_etal91} Falgarone, E.,
Phillips, T.~G., \& Walker, C.~K.\ 1991, \apj, 378, 186


\bibitem[Franco et al.(1994)]{Franco_etal94} Franco, J., Shore, S.~N.,
\& Tenorio-Tagle, G.\ 1994, \apj, 436, 795

\bibitem[Fuller \& Myers(1992)]{Fuller_Myers92} Fuller, G.~A., \&
Myers, P.~C.\ 1992, \apj, 384, 523



 \bibitem[Gazol, V\'azquez-Semadeni \& Kim(2005)]{Gazol_etal05} Gazol,
 A., V{\'a}zquez-Semadeni, E., \& Kim, J.\ 2005, \apj, 630, 911

\bibitem[Goldreich \& Kwan(1974)]{Goldreich_Kwan74} Goldreich, P., \&
Kwan, J.\ 1974, \apj, 189, 441

\bibitem[Hartmann(1998)]{Hartmann98} Hartmann, L.\ 1998, Accretion
processes in star formation.  Cambridge University Press
(Cambridge:New York)


\bibitem[Hartmann et al.(2001)]{HBB01} Hartmann, L.,
Ballesteros-Paredes, J., \& Bergin, E.~A.\ 2001, \apj, 562, 852

\bibitem[Hartmann \& Burton(1997)]{Hartmann_Burton97} Hartmann, D., \&
Burton, W.~B.\ 1997, Atlas of Galactic Neutral Hydrogen.  Cambridge
University Press



\bibitem[Heitsch et~al. (2001)]{HMK01} {Heitsch}, F., {Mac Low},
M.-M., and {Klessen}, R.~S. 2001, \apj, 547, 280--291

\bibitem[Heyer et al.(2001)]{Heyer_etal01} Heyer, M.~H., Carpenter, 
J.~M., \& Snell, R.~L.\ 2001, \apj, 551, 852 


\bibitem[Hunter \& Fleck(1982)]{Hunter_Fleck82} Hunter, J.~H., Jr., \&
Fleck, R.~C., Jr.\ 1982, \apj, 256, 505


 \bibitem[Jenkins(2002)]{Jenkins02} Jenkins, E.~B.\ 2002, \apj, 580,
 938

 \bibitem[Jenkins et al.(1983)]{Jenkins_etal83} Jenkins, E.~B., Jura,
 M., \& Loewenstein, M.\ 1983, \apj, 270, 88

 \bibitem[Jenkins \& Tripp(2001)]{Jenkins_Tripp01} Jenkins, E.~B., \&
 Tripp, T.~M.\ 2001, \apjs, 137, 297




\bibitem[Jijina et al.(1999)]{Jijina_etal99} Jijina, J., Myers, P.~C.,
\& Adams, F.~C.\ 1999, \apjs, 125, 161

\bibitem[Kegel(1989)]{Kegel89} Kegel, W.~H.\ 1989, \aap, 225, 517

\bibitem[Keto \& Myers(1986)]{Keto_Myers86} Keto, E.~R., \& Myers, 
P.~C.\ 1986, \apj, 304, 466 


\bibitem[Klessen et~al.(2000)]{KHM00} {Klessen}, R.~S., {Heitsch}, F.,
and {Mac Low}, M.-M. (2000) \apj, 535, 887--906

\bibitem[Kolmogorov(1941)]{Kolmogorov41} {Kolmogorov}, A.~N. 1941,
Dokl. Akad. Nauk SSSR, 30, 301--305

\bibitem[Kornreich \& Scalo(2000)]{Kornreich_Scalo00} Kornreich, P.,
\& Scalo, J.\ 2000, \apj, 531, 366

\bibitem[Krumholz \& McKee(2005)]{Krumholz_McKee05} Krumholz, M.~R., \& 
McKee, C.~F.\ 2005, \apj, 630, 250 


\bibitem[Larson(1981)]{Larson81} {Larson}, R.~B. 1981, \mnras 194,
809--826

\bibitem[Lesieur(1990)]{Lesieur90} Lesieur, M. 1990. Turbulence in
 Fluids (Dordrecht:Kluwer)


\bibitem[Lequeux(2005)]{Lequeux05} Lequeux, J.\ 2005, The interstellar
medium.  2005 EDP Sciences, Astronomy and astrophysics library,
(Berlin: Springer)


\bibitem[Loren(1989)]{Loren89} Loren, R.~B.\ 1989, \apj, 338, 
925 

\bibitem[Mac Low et al.(2005)]{MacLow_etal05} Mac Low, M.-M.,
 Balsara, D.~S., Kim, J., \& de Avillez, M.~A.\ 2005, \apj, 626, 864
 

\bibitem[Mac Low and Klessen(2004)]{MK04} {Mac Low}, M.-M. and
{Klessen}, R.~S. 2004, Rev.~Mod.~Phys., 76, 125--194


\bibitem[McCray \& Snow(1979)]{McCray_Snow79} McCray, R., \& Snow,
T.~P., Jr.\ 1979, \araa, 17, 213

\bibitem[McCrea(1957)]{McCrea57} McCrea, W.~H.\ 1957, \mnras, 
117, 562 


\bibitem[McKee(1999)]{McKee99} McKee, C.~F.\ 1999, NATO ASIC 
Proc.~540: The Origin of Stars and Planetary Systems, 29 




\bibitem[McKee and Zweibel(1992)]{MZ92} {McKee}, C.~F. and {Zweibel},
E.~G. 1992, \apj 399, 551--562


\bibitem[Mellema et al.(2005)]{Mellema_etal06} Mellema, G., Arthur,
S.~J., Henney, W.~J., Iliev, I.~T., \& Shapiro, P.~R.\ 2005, ArXiv
Astrophysics e-prints, arXiv:astro-ph/0512554


\bibitem[Moriarty-Schieven et al.(1997)]{Moriarty-Schieven_etal97}
Moriarty-Schieven, G.~H., Andersson, B.-G., \& Wannier, P.~G.\ 1997,
\apj, 475, 642

\bibitem[Myers \& Goodman(1988)]{MG88} Myers, P.~C., \& Goodman,
A.~A.\ 1988, \apjl, 326, L27

\bibitem[Nakano(1998)]{Nakano98} Nakano, T.\ 1998, \apj, 494, 587


\bibitem[Padoan(1995)]{Padoan95} {Padoan}, P. 1995, \mnras 277,
377--388

\bibitem[Padoan \& Nordlund(1999)]{PN99} Padoan, P., \& Nordlund,
{\AA}.\ 1999, \apj, 526, 279


\bibitem[Parker(1969)]{Parker69} Parker, E.~N.\ 1969, Space Science
Reviews, 9, 651

\bibitem[Parker(1979)]{Parker79} Parker, E.~N.\ 1979.  Cosmical
magnetic fields: Their origin and their activity.  Oxford University
Press (Oxford, Clarendon)

\bibitem[Passot et al.(1995)]{PVP95} Passot, T., Vazquez-Semadeni, E.,
\& Pouquet, A.\ 1995, \apj, 455, 536


 \bibitem[Redfield \& Linsky(2004)]{Redfield_Linsky04} Redfield, S.,
 \& Linsky, J.~L.\ 2004, \apj, 613, 1004

\bibitem[Sasao(1973)]{Sasao73} {Sasao}, T. 1973, Publ.  Astron.  Soc.
Jap., 25, 1--33

\bibitem[Scalo(1988)]{Scalo88} Scalo, J.~M.\ 1988, LNP Vol.~315:
Molecular Clouds, MilkY-Way and External Galaxies, 315, 201


\bibitem[Scalo(1990)]{Scalo90} Scalo, J.\ 1990, ASSL Vol.~162:
Physical Processes in Fragmentation and Star Formation, 151

\bibitem[Shadmehri et al.(2002)]{Shadmehri_etal02} Shadmehri, M.,
V{\'a}zquez-Semadeni, E., \& Ballesteros-Paredes, J.\ 2002, ASP
Conf.~Ser.~276: Seeing Through the Dust: The Detection of HI and the
Exploration of the ISM in Galaxies, 276, 190

\bibitem[Shu(1991)]{Shu91} Shu, F.\ 1991.  The Physics of the
Astrophysics. II. Gas dynamics. Published by University Science Books,
(New York)


\bibitem[Spitzer(1978)]{Spitzer78} Spitzer, L.\ 1978.  Physical
processes in the Interstellar Medium. New York Wiley-Interscience

\bibitem[Stahler \& Palla(2005)]{Stahler_Palla05} Stahler, S.~W., \&
Palla, F.\ 2005, The Formation of Stars, by Steven W.~Stahler,
Francesco Palla, Wiley-VCH

\bibitem[Tan et al.(2006)]{Tan_etal06} Tan, J.~C., Krumholz, 
M.~R., \& McKee, C.~F.\ 2006, \apjl, 641, L121

\bibitem[Tilley \& Pudritz(2004)]{TP04} Tilley, D.~A., \& Pudritz,
R.~E.\ 2004, \mnras, 353, 769

\bibitem[Vazquez-Semadeni et al.(1997)]{VBR97} Vazquez-Semadeni, E.,
Ballesteros-Paredes, J., \& Rodriguez, L.~F.\ 1997, \apj, 474, 292

\bibitem[V{\'a}zquez-Semadeni et al.(2003)]{VBK03}
V{\'a}zquez-Semadeni, E., Ballesteros-Paredes, J., \& Klessen, R.~S.\
2003, \apjl, 585, L131

\bibitem[Vazquez-Semadeni \& Gazol(1995)]{VSG95} Vazquez-Semadeni, E.,
\& Gazol, A.\ 1995, \aap, 303, 204

 \bibitem[V{\'a}zquez-Semadeni et al(2003)]{VS_etal03}
 V{\'a}zquez-Semadeni, E., Gazol, A., Passot, T., \&
 S\'anchez-Salcedo, J. \ 2003, LNP Vol.~614: Turbulence and Magnetic
 Fields in Astrophysics, 614, 213

\bibitem[V{\'a}zquez-Semadeni et al.(2005)]{VKB05}
V{\'a}zquez-Semadeni, E., Kim, J., \& Ballesteros-Paredes, J.\ 2005,
\apjl, 630, L49


\bibitem[V\'azquez-Semadeni et al.(1996)]{VPP96} V\'azquez-Semadeni,
E., Passot, T., \& Pouquet, A.\ 1996, \apj, 473, 881


\bibitem[von Weizs{\" a}cker(1951)]{VonWeiczacker51} {von Weizs{\"
a}cker}, C.~F. 1951, \apj, 114, 165--186.


\bibitem[Ward-Thompson et al.(2006)]{Ward-Thompson_etal06} Ward-Thompson,
D., Andre, P., Crutcher, R., Johnstone, D., Onishi, T., \& Wilson, C.\
2006, (astro-ph/0603474)


\bibitem[Williams \& Maddalena(1996)]{Williams_Maddalena96} Williams,
J.~P., \& Maddalena, R.~J.\ 1996, \apj, 464, 247


\bibitem[Yonekura et al.(1997)]{Yonekura_etal97} Yonekura, Y., Dobashi, 
K., Mizuno, A., Ogawa, H., \& Fukui, Y.\ 1997, \apjs, 110, 21 



\bibitem[Zuckerman \& Evans(1974)]{Zuckerman_Evans74} Zuckerman, B.,
\& Evans, N.~J., II 1974, \apjl, 192, L149

\end{thebibliography}
 \end{document}